\documentclass[a4paper,11pt]{article}
\usepackage{a4wide,amsmath,amssymb,bbm}
\usepackage[english]{babel}
\usepackage[utf8]{inputenc}

\usepackage{color}

\usepackage{hyperref,enumerate}
\hypersetup{
    colorlinks,
    linkcolor={black},
    citecolor={black},
    urlcolor={black}
}

\DeclareMathOperator{\tr}{tr}

\makeatletter

\@addtoreset{equation}{section} \makeatother

\begin{document}

\hfill
\vspace{30pt}

\begin{center}
{\huge{\bf Relaxing unimodularity for Yang-Baxter deformed strings}}

\vspace{80pt}

Stanislav Hronek  \ \ and \ \ Linus Wulff

\vspace{15pt}

\small {\it Department of Theoretical Physics and Astrophysics, Faculty of Science, Masaryk University\\ 611 37 Brno, Czech Republic}
\\
\vspace{12pt}
\texttt{436691@mail.muni.cz, wulff@physics.muni.cz}\\

\vspace{100pt}

{\bf Abstract}
\end{center}
\noindent
We consider so-called Yang-Baxter deformations of bosonic string sigma-models, based on an $R$-matrix solving the (modified) classical Yang-Baxter equation. It is known that a unimodularity condition on $R$ is sufficient for Weyl invariance at least to two loops (first order in $\alpha'$). Here we ask what the necessary condition is. We find that in cases where the matrix $(G+B)_{mn}$, constructed from the metric and $B$-field of the undeformed background, is degenerate the unimodularity condition arising at one loop can be replaced by weaker conditions. We further show that for non-unimodular deformations satisfying the one-loop conditions the Weyl invariance extends at least to two loops (first order in $\alpha'$). The calculations are simplified by working in an $O(D,D)$-covariant doubled formulation.

\clearpage
\setcounter{page}{1}
\tableofcontents


\section{Introduction}
Yang-Baxter (YB) deformations of 2D $\sigma$-models were introduced by Klim\v c\'ik in \cite{Klimcik:2002zj}. The name comes from the fact that the deformation is constructed using an $R$-matrix which solves the (modified) classical Yang-Baxter equation. It was later realized that these deformations preserve the classical integrability of the $\sigma$-model \cite{Klimcik:2008eq}. Delduc, Magro and Vicedo constructed the YB deformation for symmetric spaces in \cite{Delduc:2013fga}, and then for the $AdS_5\times S^5$ superstring in \cite{Delduc:2013qra}, based on the Drinfeld-Jimbo $R$-matrix solving the modified classical YB equation. Shortly thereafter it was shown in \cite{Kawaguchi:2014qwa} that essentially the same construction works also for $R$-matrices solving the ordinary (non-modified) classical YB equation. The latter are often referred to as homogenous YB deformations and have an interesting realization in terms of non-abelian T-duality \cite{Hoare:2016wsk,Borsato:2016pas,Borsato:2017qsx}.

Surprisingly it was found, starting with the paper \cite{Arutyunov:2015qva}, that the backgrounds corresponding to these deformed string $\sigma$-models did not always satisfy the equations of supergravity, but a certain generalization of these \cite{Arutyunov:2015mqj,Wulff:2016tju}. When this is the case the deformed $\sigma$-model is only scale invariant, but not Weyl invariant, at one loop and cannot be interpreted as a consistent string. For supercoset models such as the $AdS_5\times S^5$ superstring a condition was found on the $R$-matrix that leads to a viable, i.e. one-loop Weyl invariant, deformed string $\sigma$-model. The $R$-matrix should be unimodular, i.e. its trace with the Lie algebra structure constants should vanish, $R^{rs}f_{rs}{}^t=0$ \cite{Borsato:2016ose}.

Subsequently, using the realization via non-abelian T-duality, homogeneous YB deformations were defined for a general Green-Schwarz superstring with isometries \cite{Borsato:2018idb}.\footnote{In the abelian case these deformations are equivalent to so-called TsT-transformations, consisting of T-duality, a coordinate shift and a T-duality back \cite{Osten:2016dvf}.} Interestingly, examples were found where a non-unimodular $R$-matrix nevertheless gave rise to a good (super)gravity background \cite{Sakamoto:2018krs,Borsato:2018spz}. Therefore, while the unimodularity condition is sufficient, it is not necessary to solve the one-loop Weyl invariance conditions, i.e. the background (super)gravity equations.

Here we will determine the precise conditions for (bosonic) YB deformations to respect one-loop Weyl-invariance. We find that, at least for deformations of symmetric spaces, the only exceptions to the unimodularity condition occur when the matrix $(G+B)_{mn}$, where $G,B$ are the metric and $B$-field of the undeformed background, is degenerate\footnote{In the homogeneous case we prove this only for rank $R<8$ for technical reasons.}.\footnote{Gauge-transformations of $B$, which could affect this, are severely restricted by the fact that $B$ is required to be invariant under the isometries involved in the deformation. This is required in the homogeneous case \cite{Borsato:2018idb}. In the inhomogeneous case with a WZ-term \cite{Delduc:2014uaa} it is not required and our analysis is incomplete in that case.} In that case, the unimodularity condition is no longer necessary and is replaced by weaker conditions which we give. This is consistent with the examples found in \cite{Sakamoto:2018krs,Borsato:2018spz} since the $AdS_3\times S^3$ background considered there has degenerate $G+B$.

We then go on to analyze what happens at two loops, i.e. when we include the first $\alpha'$-correction to the (super)gravity equations. We find that the conditions at two loops are weaker and only a subset of the one-loop conditions are needed.

These calculations are simplified enormously by working with the $O(D,D)$-covariant formulation known as Double Field Theory (DFT). In DFT a manifestly $O(D,D)$-covariant formulation is achieved by doubling the coordinates to $X^M=(\tilde x_m,x^m)$. One then imposes a "section condition" which effectively removes half of them, leaving the right number of physical coordinates. Here we will work only with the standard choice of section, $X^M=(0,x^m)$ or $\partial_M=(0,\partial_m)$, and therefore the coordinates are not doubled. However, the tangent space is effectively doubled and there are two copies of the Lorentz group. Therefore there are two sets of vielbeins $e^{(+)}$ and $e^{(-)}$ which transform independently under each Lorentz group factor. Fixing the gauge $e^{(+)}=e^{(-)}=e$ breaks the doubled Lorentz group down to its diagonal, which becomes the standard Lorentz group. With this gauge fixing the action and equations of motion of the doubled formulation reduce to those of standard (super)gravity. The reason the doubled formulation is useful is that the YB deformation becomes equivalent to a coordinate dependent $O(D,D)$-transformation which is easy to analyze. In fact the so-called generalized fluxes, the basic fields of the so-called flux formulation we are using \cite{Geissbuhler:2013uka}, transform very simply under the YB deformation. The 3-form flux is invariant while the 1-form acquires a shift. This shift vanishes in the unimodular case and the generalized fluxes are simply invariant, from which one can immediately conclude that such YB deformations preserve Weyl invariance at least to two loops \cite{Borsato:2020bqo}. In the present case, we are interested in non-unimodular $R$-matrices and we have to take the shift into account. Provided that this shift satisfies certain conditions, which we determine, the Weyl-invariance is preserved at least up to two loops. It is interesting that it is possible to shift the 1-form generalized flux in certain ways and still preserve the equations of the doubled formulation including the first $\alpha'$-correction. This should have an interpretation in gauged DFT \cite{Grana:2012rr}, but we will not pursue this here.

In \cite{Borsato:2020bqo} the doubled formulation was used to determine the first $\alpha'$-correction to the deformed background for unimodular $R$. This correction arises because the fields of the doubled formulation are not Lorentz-covariant once $\alpha'$-corrections are included and a double Lorentz transformation is needed to go to the gauge $e^{(+)}=e^{(-)}=e$ and reduce to the standard (super)gravity fields, thus leading to a correction to the background.\footnote{The correction agrees with what is found by a much more involved calculation using standard (super)gravity \cite{Borsato:2019oip}.} Our analysis here shows that no additional corrections are needed in the non-unimodular case, so the correction to the deformed background is still given by the expressions found in \cite{Borsato:2020bqo}.

The outline of this paper is as follows. First we review the elements we need of the flux formulation of DFT and how the $\alpha'$-correction to the double Lorentz transformations determine the action to the first order in $\alpha'$. In section \ref{sec:YB} we derive the conditions for a YB deformation to lead to a (super)gravity background, i.e. the conditions needed for one-loop Weyl-invariance. The situation at two loops is analyzed in section \ref{sec:2loop} where we find weaker conditions than at one loop. We end with some conclusions.

\section{Doubled (flux) formulation}
The $O(D,D)$-covariant formulation of (super)gravity used in DFT \cite{Aldazabal:2013sca,Hohm:2013bwa,Berman:2013eva} turns out to be very powerful for the kinds of questions we are interested in here. In particular we will work with a frame-like formulation of DFT \cite{Siegel:1993th,Siegel:1993xq,Hohm:2010xe} where the structure group consists of two copies of the Lorentz group $O(1,D-1)\times O(D-1,1)$. In particular we use the so-called flux formulation of \cite{Geissbuhler:2013uka,Marques:2015vua} where the first $\alpha'$-correction to the bosonic and heterotic string can also be nicely incorporated. We will always assume that the section condition is solved in the standard way $\partial_M=(0,\partial_m)$ so that we are really just working with a rewriting of (super)gravity.

The starting point is to introduce a generalized (inverse) vielbein parametrized as
\begin{equation}
E_A{}^M=
\frac{1}{\sqrt2}
\left(
\begin{array}{cc}
e^{(+)a}{}_m-e^{(+)an}B_{nm} & e^{(+)am}\\
-e^{(-)}_{am}-e^{(-)}_a{}^nB_{nm} & e^{(-)}_a{}^m
\end{array}
\right)\,.
\label{eq:E}
\end{equation}
Here $e^{(\pm)}$ are two sets of vielbeins for the metric $G_{mn}$ which transform independently as $\Lambda^{(\pm)}e^{(\pm)}$ under the two Lorentz-group factors. To go to the standard supergravity picture one fixes a gauge $e^{(+)}=e^{(-)}=e$, leaving only one copy of the Lorentz-group. The dilaton $\Phi$ is encoded in the generalized dilaton $d$ defined as
\begin{equation}
e^{-2d}=e^{-2\Phi}\sqrt{-G}\,.
\end{equation}
There are two constant metrics, the $O(D,D)$-metric $\eta^{AB}$ and the generalized metric $\mathcal H^{AB}$ which take the form
\begin{equation}
\eta^{AB}=\eta_{AB}=
\left(
\begin{array}{cc}
	\bar\eta & 0\\
	0 & -\bar\eta
\end{array}
\right)\,,\qquad
\mathcal H^{AB}=
\left(
\begin{array}{cc}
	\bar\eta & 0\\
	0 & \bar\eta
\end{array}
\right)\,,
\end{equation}
where $\bar\eta=(-1,1,\ldots,1)$ is the usual Minkowski metric. The flat tangent space indices $A,B,\ldots$ are raised and lowered with $\eta^{AB}$, $\eta_{AB}$. The generalized vielbein is used to convert between these indices and coordinate indices $M,N,\ldots$. In particular we have the usual expressions for the $O(D,D)$-metric and the generalized metric in a coordinate basis
\begin{align}
\eta^{MN}=&
E_A{}^M\eta^{AB}E_B{}^N=
\left(
\begin{array}{cc}
	0 & \delta_m{}^n\\
	\delta^m{}_n & 0
\end{array}
\right)\,,\\
\mathcal H^{MN}=&E_A{}^M\mathcal H^{AB}E_B{}^N=
\left(
\begin{array}{cc}
	G_{mn}-B_{mk}G^{kl}B_{ln} & B_{mk}G^{kn}\\
	-G^{mk}B_{kn} & G^{mn}
\end{array}
\right)\,.
\end{align}
We also define
\begin{equation}
\partial_A=E_A{}^M\partial_M\,,
\end{equation}
where $\partial_M=(0,\partial_m)$ is the ordinary derivative.

The basic fields of the flux formulation are the generalized fluxes. These are constructed from the generalized vielbein as
\begin{equation}
\mathcal F_{ABC}=3\partial_{[A}E_B{}^ME_{C]M}\,,\qquad\mathcal F_A=\partial^BE_B{}^ME_{AM}+2\partial_Ad\,.
\label{eq:fluxes}
\end{equation}
The importance of these objects comes from the fact that they transform as scalars under generalized diffeomorphisms implemented by the generalized Lie derivative defined as
\begin{equation}
\mathcal L_XY^M=X^N\partial_NY^M+(\partial^MX_N-\partial_NX^M)Y^N\,.
\label{eq:genLie}
\end{equation}
The generalized diffeomorphisms contain the usual diffeomorphisms and $B$-field gauge-transformations. The generalized fluxes satisfy the following Bianchi identities
\begin{equation}
4\partial_{[A}\mathcal F_{BCD]}=3\mathcal F_{[AB}{}^E\mathcal F_{CD]E}\,,\qquad
2\partial_{[A}\mathcal F_{B]}=-(\partial^C-\mathcal F^C)\mathcal F_{ABC}\,.
\label{eq:Bianchi}
\end{equation}
Note also that
\begin{equation}
[\partial_A,\partial_B]=\mathcal F_{ABC}\,\partial^C\,.
\label{eq:d-comm}
\end{equation}

The bosonic/heterotic\footnote{Setting the gauge fields and fermions of the heterotic string to zero.} string low-energy effective action can be cast in doubled form as
\begin{equation}
S=\int dX\,e^{-2d}\mathcal R\,,
\label{eq:S0}
\end{equation}
where the generalized Ricci scalar is defined as\footnote{The last two terms are often written instead as $\frac14\mathcal F_{ACD}\mathcal F_B{}^{CD}\mathcal H^{AB}-\frac{1}{12}\mathcal F_{ABC}\mathcal F_{DEF}\mathcal H^{AD}\mathcal H^{BE}\mathcal H^{CF}-\frac16\mathcal F_{ABC}\mathcal F^{ABC}$. In terms of the generalized metric we have instead 
$$
\mathcal R
=
4\partial_M(\mathcal H^{MN}\partial_Nd)
-\partial_M\partial_N\mathcal H^{MN}
-4\mathcal H^{MN}\partial_Md\partial_Nd
+\frac18\mathcal H^{MN}\partial_M\mathcal H^{KL}\partial_N\mathcal H_{KL}
-\frac12\mathcal H^{MN}\partial_M\mathcal H^{KL}\partial_K\mathcal H_{LN}\,.
$$
}
\begin{equation}
\mathcal R
=
-4\partial^A\mathcal F_A^{(-)}
+2\mathcal F^A\mathcal F_A^{(-)}
-\mathcal F^{(-)}_{ABC}\mathcal F^{(-)ABC}
-\frac13\mathcal F^{(--)}_{ABC}\mathcal F^{(--)ABC}\,.
\label{eq:R}
\end{equation}
Here we have defined certain projections of the generalized fluxes using the natural projection operators
\begin{equation}
P_\pm=\frac12\left(\eta\pm\mathcal H\right)\,,
\end{equation}
as follows
\begin{equation}
\mathcal F^{(\pm)}_A=(P_\pm)_A{}^B\mathcal F_B\,,
\end{equation}
and
\begin{equation}
\mathcal F^{(\pm)}_{ABC}=(P_{\mp})_A{}^D(P_{\pm})_B{}^E(P_{\pm})_C{}^F\mathcal F_{DEF}\,,\qquad
\mathcal F^{(\pm\pm)}_{ABC}=(P_{\pm})_A{}^D(P_{\pm})_B{}^E(P_{\pm})_C{}^F\mathcal F_{DEF}\,.
\end{equation}
Setting $e^{(+)}=e^{(-)}=e$ in the generalized vielbein (\ref{eq:E}) this can be shown to reduce to the correct low-energy effective (super)gravity action.

We will be interested in whether certain transformations of the generalized fluxes map a solution to another solution, so we will need the equations of motion following from the action (\ref{eq:S0}). These can be easily found using the variations of the generalized fluxes with respect to the generalized vielbein and dilaton
\begin{equation}
\delta_E\mathcal F_{ABC}=3\partial_{[A}\delta E_{BC]}+3\delta E_{[A}{}^D\mathcal F_{BC]D}\,,\quad
\delta_E\mathcal F_A=\partial^B\delta E_{BA}+\delta E_A{}^B\mathcal F_B\,,\quad
\delta_d\mathcal F_A=2\partial_A\delta d\,,
\label{eq:deltaF}
\end{equation}
where $\delta E_{AB}=\delta E_A{}^ME_{BM}$ is anti-symmetric by construction. The equations of motion become
\begin{align}
\mathcal R=0\,,\qquad\partial^{(+)}_A\mathcal F^{(-)}_B+(\partial^C-\mathcal F^C)\mathcal F^{(-)}_{ABC}-\mathcal F^{(+)}_{CDA}\mathcal F^{(-)DC}{}_B=0\,.
\label{eq:eom}
\end{align}
Here we have defined the projected derivatives $\partial^{(\pm)}_A=(P_\pm)_A{}^B\partial_B$. The second equation of motion can equivalently be written with the opposite projections by exchanging + and - superscripts. Setting $e^{(+)}=e^{(-)}=e$ they reduce to correct (super)gravity equations of motion.

The action (\ref{eq:S0}) is invariant under three important symmetries. The first is generalized diffeomorphisms, which encode regular diffeomorphisms and $B$-field gauge transformations. In the flux formulation we are working with here the generalized diffeomorphism invariance is manifest since the fluxes and the derivative $\partial_A$ transform as scalars. The second symmetry is that of global $O(D,D)$-transformations
\begin{equation}
X^M\rightarrow X^Nh_N{}^M\,,\qquad E_A{}^M\rightarrow E_A{}^Nh_N{}^M\qquad\mbox{with}\qquad h_M{}^N\in O(D,D)
\end{equation}
and $h_M{}^N$ constant. Again the action is manifestly invariant under these transformations since the fluxes are invariant. In our case we are always imposing the standard section condition $\partial_M=(0,\partial_m)$ so this symmetry is (partially) broken.

Finally, the most important symmetry for us will be the invariance under double Lorentz transformations
\begin{equation}
\delta E_A{}^ME_{BM}=\delta E_{AB}=\lambda_{AB}\qquad\mbox{with}\qquad (P_+)_A{}^C(P_-)_B{}^D\lambda_{CD}=0\,.
\end{equation}
The parameters of the infinitesimal double Lorentz transformation $\lambda_{AB}$ commute with the projectors $P_\pm$ so their non-trivial components are $\lambda^{(+)}_{AB}$ and $\lambda^{(-)}_{AB}$, corresponding to the two copies of the Lorentz group. These two copies rotate the two vielbeins $e^{(\pm)}$ in (\ref{eq:E}) independently. The (double) Lorentz invariance of the action (\ref{eq:S0}) is not manifest. It can be verified with a bit of algebra using the variations of the fluxes (\ref{eq:deltaF}). In particular it follows from these expressions that under a double Lorentz transformation
\begin{equation}
\delta\mathcal F^{(\pm)}_{ABC}=\lambda^{(\mp)}_A{}^D\mathcal F^{(\pm)}_{DBC}+2\lambda^{(\pm)}_{[B}{}^D\mathcal F^{(\pm)}_{|AD|C]}+\partial^{(\mp)}_A\lambda^{(\pm)}_{BC}\,,
\end{equation}
which, except for the projections, is precisely the transformation of a connection. Indeed, suppressing the last two indices we have\footnote{We will try to be clear about when we suppress the last two indices to avoid possible confusion with the generalized flux with one index $\mathcal F_A$.}
\begin{equation}
\delta\mathcal F^{(\pm)}_M=\partial^{(\mp)}_M\lambda^{(\pm)}+[\lambda^{(\pm)},\mathcal F^{(\pm)}_M]\,,
\label{eq:delta-Fpm}
\end{equation}
so that $\mathcal F^{(\pm)}_M$ behave very much like connections. In fact, fixing the double Lorentz transformations by setting $e^{(+)}=e^{(-)}=e$ the non-zero components of $\mathcal F^{(\pm)}$ are \cite{Marques:2015vua}
\begin{equation}
\mathcal F^{(+)}_M{}^{ab}=
\frac12\left(
\begin{array}{c}
G^{mn}\omega^{(+)ab}_n\\
-(1-BG)_m{}^n\omega^{(+)ab}_n
\end{array}
\right)\,,
\qquad
\mathcal F^{(-)}_{Mab}=
\frac12\left(
\begin{array}{c}
G^{mn}\omega^{(-)}_{nab}\\
(1+BG)_m{}^n\omega^{(-)}_{nab}
\end{array}
\right)\,,
\label{eq:Fpm}
\end{equation}
where $\omega_m^{(\pm)cd}=\omega_m{}^{cd}\pm\frac12H_m{}^{cd}$. These expressions will be useful later.

A very important point is that the double Lorentz transformations receive $\alpha'$-corrections. In fact, this is a good thing since it allows us to derive the first $\alpha'$-correction to the action (\ref{eq:S0}) from the knowledge of the correction to the Lorentz transformations. We will now see how this works.

\subsection{The first \texorpdfstring{$\alpha'$}{alpha'}-correction}
At the first order in $\alpha'$ the double Lorentz transformations get corrected to \cite{Marques:2015vua}
\begin{equation}
\delta E_{AB}=\lambda_{AB}
+a\tr\big(\partial_{[A}^{(-)}\lambda\mathcal F^{(-)}_{B]}\big)
-b\tr\big(\partial_{[A}^{(+)}\lambda\mathcal F^{(+)}_{B]}\big)
\,,
\label{eq:deltaE}
\end{equation}
where $a=b=-\alpha'$ for the bosonic string and $a=-\alpha'$, $b=0$ for the heterotic string ($a=b=0$ for type II). The correction involves the connection-like objects $\mathcal F^{(\pm)}_{ABC}$ (note the trace over the last two indices) and is therefore of the form of a Green-Schwarz transformation.

The knowledge of the correction to the Lorentz transformation can be used to find the $\alpha'$-correction to the action \cite{Marques:2015vua}, as we will now review. For simplicity we will set $b=0$ in the derivation and restore $b$ at the end. The variation (\ref{eq:deltaE}) is then of the form $\delta=\delta^0+a\delta^1$ and a short calculation gives for the projections of the generalized fluxes appearing in the lowest order action (\ref{eq:S0}), (\ref{eq:R})
\begin{equation}
\delta^1\mathcal F_A^{(-)}=-\frac12(\partial^B-\mathcal F^B)\tr\big(\partial_A^{(-)}\lambda\mathcal F^{(-)}_B\big)\,,
\label{eq:d1FA}
\end{equation}
\begin{equation}
\delta^1\mathcal F^{(--)}_{ABC}=\frac32\tr\big(\partial_{[A}^{(-)}\lambda\mathcal F^{(-)D}\big)\mathcal F^{(-)}_{D|BC]}
\label{eq:d1F--}
\end{equation}
and
\begin{equation}
\delta^1\mathcal F^{(-)}_{ABC}=
(P_-)_{[C}{}^D\tr\big(\partial^{(-)}_{B]}\lambda\mathcal R^{(-)}_{AD}\big)
+\frac12\tr\big(\mathcal F^{(-)}_A\partial^D\lambda\big)\mathcal F^{(-)}_{DBC}
+\frac12\tr\big(\partial_{[B}^{(-)}\lambda\mathcal F^{(-)D}\big)\mathcal F^{(+)}_{C]AD}\,.
\label{eq:d1F-}
\end{equation}
In the last expression we have defined the 'curvature' of the 'connection' $\mathcal F^{(-)}_{ABC}$ as (suppressing the last two indices which are projected by $P_-$)
\begin{equation}
\mathcal R^{(-)}_{AB}
=
2\partial^{\phantom{(-)}}_{[A}\mathcal F^{(-)}_{B]}
-(P_+)_{[B}{}^D\mathcal F_{A]DE}\mathcal F^{(-)E}
-[\mathcal F^{(-)}_A,\mathcal F^{(-)}_B]\,.
\label{eq:RAB}
\end{equation}
This object will be useful later. In particular when we project the indices $A$ and $B$ with $P_+$ we have, writing $\bar{\mathcal R}^{(-)}_{AB}=(P_+)_A{}^C(P_+)_B{}^D\mathcal R^{(-)}_{CD}$, (again the last two indices are suppressed)
\begin{equation}
\delta^0\bar{\mathcal R}^{(-)}_{AB}
=
2\lambda^{(+)}_{[A}{}^C\bar{\mathcal R}^{(-)}_{|C|B]}
+[\lambda^{(-)},\mathcal R^{(-)}_{AB}]
+\mathcal F^{(+)}_{CAB}\partial^C\lambda^{(-)}
-\partial^C\lambda^{(+)}_{AB}\mathcal F^{(-)}_C\,,
\label{eq:dRAB}
\end{equation}
which apart from the last two terms is the expected transformation of a curvature.

At lowest order in $\alpha'$ the action is Lorentz invariant. At the next order we find
\begin{equation}
\delta^1\mathcal R
=
-4(\partial^A-\mathcal F^A)\delta^1\mathcal F_A^{(-)}
-2\partial^B\mathcal F^A\tr\big(\partial_A^{(-)}\lambda\mathcal F^{(-)}_B\big)
-\frac23\mathcal F^{(--)ABC}\delta^1\mathcal F^{(--)}_{ABC}
-2\mathcal F^{(-)ABC}\delta^1\mathcal F^{(-)}_{ABC}\,.
\end{equation}
Using the expressions for the $\delta^1$-variations (\ref{eq:d1FA}), (\ref{eq:d1F--}) and (\ref{eq:d1F-}) as well as the Bianchi identity for $\mathcal F_A$ (\ref{eq:Bianchi}), (\ref{eq:d-comm}) and the section condition this becomes
\begin{align}
\delta^1\mathcal R
=&
\delta^0\left(-\partial^A\big[(\partial^B-\mathcal F^B)\tr\big(\mathcal F_A^{(-)}\mathcal F^{(-)}_B\big)\big]+(\partial^A-\mathcal F^A)\big[\mathcal F^B\tr\big(\mathcal F_A^{(-)}\mathcal F^{(-)}_B\big)\big]\right)
\nonumber\\
&{}
-\mathcal F^{ABC}\tr\big(\partial_A\partial_B^{(+)}\lambda\mathcal F^{(-)}_C\big)
+\mathcal F^{ABC}\tr\big(\partial_A^{(+)}\lambda\partial_B\mathcal F^{(-)}_C\big)
-2\mathcal F^{ABC}\tr\big(\partial_C\lambda\partial_A\mathcal F^{(-)}_B\big)
\nonumber\\
&{}
+2\mathcal F^{(-)ABC}\tr\big(\partial_B\lambda\mathcal R^{(-)}_{CA}\big)
+\mathcal F^{ABC}\partial_C\lambda_{AD}\tr\big(\mathcal F^{(-)D}\mathcal F^{(-)}_B\big)
+\partial^B\lambda^{AC}\partial_A\tr\big(\mathcal F_C^{(-)}\mathcal F^{(-)}_B\big)
\nonumber\\
&{}
+\mathcal F^{ABC}\mathcal F_{BCD}\tr\big(\partial^D\lambda\mathcal F^{(-)}_A\big)
-\mathcal F^{(--)ABC}\mathcal F^{(-)}_{DBC}\tr\big(\partial_A\lambda\mathcal F^{(-)D}\big)
\nonumber\\
&{}
-\mathcal F^{(-)ABC}\mathcal F^{(-)}_{DBC}\tr\big(\partial^D\lambda\mathcal F^{(-)}_A\big)
-\mathcal F^{(-)ABC}\mathcal F^{(+)}_{CAD}\tr\big(\partial_B\lambda\mathcal F^{(-)D}\big)\,.
\end{align}
We must now find terms of order $\alpha'$ whose lowest order Lorentz transformation cancels the terms on the RHS. The first term on the second line must be canceled by the variation of a term of the form $\mathcal F^{ABC}\tr\big(\partial_A\mathcal F_B^{(-)}\mathcal F^{(-)}_C\big)$ and we find
\begin{align}
\delta^1\mathcal R
=&
\delta^0\left(-\partial^A\big[(\partial^B-\mathcal F^B)\tr\big(\mathcal F_A^{(-)}\mathcal F^{(-)}_B\big)\big]+(\partial^A-\mathcal F^A)\big[\mathcal F^B\tr\big(\mathcal F_A^{(-)}\mathcal F^{(-)}_B\big)\big]\right)
\nonumber\\
&{}
-\delta^0\left[\mathcal F^{ABC}\tr\big(\partial_A\mathcal F_B^{(-)}\mathcal F^{(-)}_C\big)\right]
-\mathcal F^{ABC}\tr\big(\partial^{(-)}_C\lambda\bar{\mathcal R}^{(-)}_{AB}\big)
+\partial^C\lambda^{AB}\tr\big(\mathcal F^{(-)}_C\bar{\mathcal R}^{(-)}_{AB}\big)
\nonumber\\
&{}
+2\partial^C\lambda^{AB}\tr\big(\mathcal F^{(-)}_A\mathcal F^{(-)}_B\mathcal F^{(-)}_C\big)
+2\mathcal F^{ABC}\tr\big(\mathcal F_A^{(-)}\mathcal F^{(-)}_B\partial^{(+)}_C\lambda\big)
\nonumber\\
&{}
+\mathcal F^{(++)}_{ABE}\partial^C\lambda^{AB}\tr\big(\mathcal F^{(-)}_C\mathcal F^{(-)E}\big)
+2\mathcal F^{ABE}\partial_B\lambda_{CA}\tr\big(\mathcal F^{(-)C}\mathcal F^{(-)}_E\big)
\nonumber\\
&{}
+\mathcal F^{ABC}\mathcal F_{DBC}\tr\big(\partial^{(+)D}\lambda\mathcal F^{(-)}_A\big)
-\mathcal F^{(-)ABC}\mathcal F^{(-)}_{DBC}\tr\big(\partial^D\lambda\mathcal F^{(-)}_A\big)\,,
\end{align}
where we used the definition of the 'curvature' in (\ref{eq:RAB}). Using (\ref{eq:dRAB}) we see that the last two terms on the second line come from the variation of $\tr\big(\bar{\mathcal R}^{(-)AB}\bar{\mathcal R}^{(-)}_{AB}\big)$ and the remaining terms are also easy to write as the variation of something. When the dust has settled one finds, reinstating $b$, that the corrected action
\begin{equation}
S=\int dX\,e^{-2d}\left(\mathcal R+a\mathcal R^{(-)}+b\mathcal R^{(+)}\right)
\label{eq:S1}
\end{equation}
is invariant under Lorentz transformations up to and including order $\alpha'$ where
\begin{align}
\mathcal R^{(-)}
=&
\partial^A\left[(\partial^B-\mathcal F^B)\tr\big(\mathcal F_A^{(-)}\mathcal F^{(-)}_B\big)\right]
-(\partial^B-\mathcal F^B)\left[\mathcal F^A\tr\big(\mathcal F_A^{(-)}\mathcal F^{(-)}_B\big)\right]
+\tfrac12\tr\big(\bar{\mathcal R}^{(-)AB}\bar{\mathcal R}^{(-)}_{AB}\big)
\nonumber\\
&{}
+\tfrac16\mathcal F^{ABC}\mathcal C^{(-)}_{ABC}\,.
\label{eq:R-}
\end{align}
In the last term we have introduced the 'Chern-Simons' form
\begin{equation}
\mathcal C^{(-)}_{ABC}=
6\tr\big(\mathcal F^{(-)}_{[A}\partial^{\phantom{(}}_B\mathcal F^{(-)}_{C]}\big)
+3(\mathcal F^{(-)}_{D[AB}-\mathcal F^{\phantom{(-)}}_{D[AB})\tr\big(\mathcal F^{(-)}_{C]}\mathcal F^{(-)D}\big)
-4\tr\big(\mathcal F^{(-)}_{[A}\mathcal F^{(-)}_B\mathcal F^{(-)}_{C]}\big)\,.
\end{equation}
The expression for $\mathcal R^{(+)}$ is obtained by reversing the projections in an obvious way. These expressions agree with the ones written in \cite{Baron:2017dvb} but are much more compact.

\section{Yang-Baxter deformations and one-loop Weyl invariance}\label{sec:YB}
Yang-Baxter deformations are closely related to a generalization of T-duality known as Poisson-Lie (PL) T-duality. In particular homogeneous YB deformations can be constructed using non-abelian T-duality \cite{Hoare:2016wsk,Borsato:2016pas}. It is therefore not surprising that they have a natural formulation in terms of DFT. In the flux formulation we are working with they are described as a coordinate dependent $O(D,D)$-transformation \cite{Araujo:2017jkb,Sakamoto:2017cpu,Borsato:2018idb}
\begin{equation}
E_A{}^M\rightarrow\tilde E_A{}^M=E_A{}^N(1+\Theta)_N{}^M\,.
\end{equation}
The only non-zero components of $\Theta_N{}^M$ are $\Theta^{mn}=k^m_rk^n_sR^{rs}$ where $k^m_r$ are Killing vectors belonging to some Lie algebra $\mathfrak g$ indexed by $(r,s,t,..)$ and $R^{rs}$ is a constant anti-symmetric matrix satisfying, in the homogeneous case, the classical YB equation
\begin{equation}
[RX,RY]-R([RX,Y]+[X,RY])=0\,,\quad\forall X,Y\in\mathfrak g\,,
\end{equation}
which implies the 'Jacobi identity' for $\Theta$\footnote{Conversely, if we don't impose any condition on $R$, this condition follows by requiring that we get a (super)gravity solution \cite{Bakhmatov:2018apn}.}
\begin{equation}
\Theta^{N[K}\partial_N\Theta^{LM]}=0\,.
\label{eq:Jacobi}
\end{equation}
If we start from a symmetric space $\sigma$-model we can also define the inhomogeneous deformation \cite{Delduc:2013fga} where $R$ satisfies the modified classical YB equation
\begin{equation}
[RX,RY]-R([RX,Y]+[X,RY])=[X,Y]\,,\quad\forall X,Y\in\mathfrak g\,.
\end{equation}
The canonical solution is the Drinfeld-Jimbo $R$-matrix defined to annihilate elements of the Cartan subalgebra and to multiply generators corresponding to positive(negative) roots by $+i(-i)$. We can define again $\Theta^{mn}=k^m_rk^n_sR^{rs}$ which also satisfies (\ref{eq:Jacobi}).\footnote{\label{foot:sym}This was first noted in special examples in \cite{Araujo:2017enj}. We thank S. van Tongeren for pointing this out to us. The fact that the RHS in the modified YB equation does not contribute can be seen as follows. For a symmetric space $\mathfrak g$ is generated by $P_a,M_{ab}$ with commutators of the form $[P,P]\sim M$, $[M,P]\sim P$ and $[M,M]\sim M$. The Killing vectors are given by $k_r{}^m=\ell_a{}^m(\hat P\mathrm{Ad}_g)^a{}_r$ (see for example \cite{Borsato:2018idb}), where $\ell_a{}^m$ are inverse vielbeins of the left-invariant one-forms and $\hat P$ projects on the Lie algebra generators $P_a$. Now since the structure constants are Ad-invariant and since they have no component corresponding to three $P_a$ generators it follows that the RHS in the modified YB equation does not contribute.}

Note that letting $R$ be multiplied by a small parameter, usually called $\eta$, these become deformations of the original background. It is not hard to show, using the definitions (\ref{eq:fluxes}), that these deformations preserve the form of the generalized fluxes up to a shift of $\mathcal F_A$ \cite{Borsato:2020bqo}
\begin{equation}
\tilde{\mathcal F}_{ABC}=\mathcal F_{ABC}\,,\qquad\tilde{\mathcal F}_A=\mathcal F_A-2K_A\,.
\label{eq:shift1}
\end{equation}
In addition derivatives of $\mathcal F_{ABC}$ are invariant, e.g. $\tilde\partial_A\tilde{\mathcal F}_{BCD}=\partial_A\mathcal F_{BCD}$. Because of the shift this is in general not true for $\mathcal F_A$, instead
\begin{equation}
\tilde\partial_A\tilde{\mathcal F}_B=\partial_A\mathcal F_B-2\partial_AK_B-2E_A{}^N\Theta_N{}^M\partial_MK_B\,.
\label{eq:shift2}
\end{equation}
The shift of $\mathcal F_A$ is given by a certain distinguished Killing vector namely $K^M=(0,K^m)$ with
\begin{equation}
K^m=\nabla_n\Theta^{mn}=\nabla_nk_r^mk_s^nR^{rs}=-\tfrac12R^{rs}f_{rs}{}^tk_t^m
\,,
\label{eq:Km}
\end{equation}
where the third step involves using the algebra of the Killing vectors. This shift vanishes precisely when $R$ is unimodular, i.e. when $R^{rs}f_{rs}{}^t=0$.\footnote{It is easy to see that the Drinfeld-Jimbo $R$-matrix of the inhomogeneous deformation is not unimodular.} In this case the generalized fluxes and their derivatives are invariant under the deformation and this directly implies that the deformation preserves Weyl-invariance at least up to order $\alpha'$ (2 loops) \cite{Borsato:2020bqo}. If we drop the unimodularity condition we will generically get a scale-invariant but not Weyl-invariant $\sigma$-model at one loop. This is reflected in the background solving the generalized supergravity equations \cite{Arutyunov:2015mqj,Wulff:2016tju} instead of the usual ones, the extra Killing vector appearing in these equations being given by $K^m$.

Here we want to ask what happens if you don't require unimodularity but still require the deformed model to preserve one-loop Weyl invariance.\footnote{This corresponds to having a solution of the generalized supergravity equations which also solves the standard supergravity equations. Such 'trivial' solutions were analyzed in \cite{Wulff:2018aku}.} We will argue that, at least in the case of symmetric spaces, it is possible to find such non-unimodular R-matrices (at least of low enough rank to be interesting) only if the combination of metric and $B$-field of the original model $G\pm B$ is a degenerate matrix. An example where this happens is for $AdS_3\times S^3$ and indeed in that case several non-unimodular R-matrices that lead to (super)gravity solutions have been found \cite{Sakamoto:2018krs,Borsato:2018spz}.

The requirement that the equations of motion (\ref{eq:eom}) remain invariant under the deformation, which is equivalent to preservation of one-loop Weyl-invariance, becomes, using (\ref{eq:shift1}) and (\ref{eq:shift2})
\begin{align}
\partial^{(+)}_AK^{(-)}_B+(P_+)_A{}^CE_C{}^N\Theta_N{}^M\partial_MK^{(-)}_B-K^C\mathcal F^{(-)}_{ABC}
=&
0\,,
\label{eq:cond1}
\\
\partial^AK^{(-)}_A
+E_A{}^N\Theta_N{}^M\partial_MK^{(-)A}
-K^A\mathcal F^{(-)}_A
+K^AK^{(-)}_A
=&
0\,.
\label{eq:cond2}
\end{align}
Since we should think of $\Theta$ as being multiplied by a small deformation parameter these equations contain terms of first and second order in this parameter (note that $K$ (\ref{eq:Km}) is of first order). These contributions then need to vanish separately.

\subsection{First order terms}
At the lowest order in the deformation we find the conditions
\begin{equation}
\partial^{(+)}_AK^{(-)}_B-K^C\mathcal F^{(-)}_{ABC}=0\,,\qquad\partial^AK^{(-)}_A-K^A\mathcal F^{(-)}_A=0\,.
\end{equation}
Using the form of the generalized vielbein (\ref{eq:E}) with $e^{(+)}=e^{(-)}=e$, the fact that $K^M=(0,K^m)$ and the form of $\mathcal F^{(-)}_{ABC}$ in (\ref{eq:Fpm}) the first equation becomes
\begin{equation}
\nabla_a[(1+B)_b{}^cK_c]-\tfrac12H_{abc}(1+B)^c{}_dK^d=0\,.
\end{equation}
Symmetrizing in $a,b$ and using the fact that $K$ is Killing we find that $\tilde K=i_KB$ is also a Killing vector. Anti-symmetrizing we find, using $\mathcal L_KB=0$, that
\begin{equation}
dK+i_{\tilde K}H=0\,.
\label{eq:dK}
\end{equation}
This equation implies that $H$ is invariant under $\tilde K$ since $\mathcal L_{\tilde K}H=di_{\tilde K}H=-ddK=0$. We also have the same equation with $K$ and $\tilde K$ exchanged since $d\tilde K=di_KB=-i_KH$ from the invariance of the $B$-field under isometries, which we have assumed here.\footnote{This seems to be required in the construction of the general homogeneous deformations \cite{Borsato:2018idb}. In the inhomogeneous case this should be relaxed \cite{Delduc:2014uaa}, but we will not try to do this here since it would take us too far afield.} From the dilaton equation we get, using the fact that $K$ and $\tilde K$ are Killing vectors, the condition
\begin{equation}
\tilde K^m\partial_m\Phi=0\,,
\end{equation}
i.e. the dilaton is invariant under the isometry generated by $\tilde K$. To summarize, the conditions we find at this order are that $\tilde K=i_KB$ generates isometries of the background fields $G,H,\Phi$ and satisfies (\ref{eq:dK}).

For our later discussion of two-loop conformal invariance it will be useful to express these conditions in terms of the generalized fluxes. The fact that $K$ and $\tilde K$ generate symmetries of the original background implies that under YB deformations
\begin{equation}
\tilde{\mathcal F}^{(\pm)}_A\tilde\partial^A(\mbox{something invariant})=\mathcal F^{(\pm)}_A\partial^A(\mbox{something invariant})\,.
\label{eq:dcond1}
\end{equation}
In addition we have
\begin{equation}
K^A\mathcal F^{(-)}_{ABC}
=
\tfrac12(K^m+\tilde K^m)\omega^{(-)}_{mbc}\delta^b_B\delta^c_C
=
-\tfrac12(\nabla_bK_c+\nabla_b\tilde K_c)\delta^b_B\delta^c_C
-\tfrac14(K^m+\tilde K^m)H_{mbc}\delta^b_B\delta^c_C
=
0\,,
\end{equation}
where we used invariance of the vielbein under $K,\tilde K$ which implies $i_K\omega_{ab}=-\nabla_aK_b$ and similarly for $\tilde K$ as well as the equation (\ref{eq:dK}) and the same with $K$ and $\tilde K$ exchanged. The same is true with the opposite projection and therefore we have
\begin{equation}
\tilde{\mathcal F}^A\tilde{\mathcal F}^{(\pm)}_{ABC}=\mathcal F^A\mathcal F^{(\pm)}_{ABC}\,.
\label{eq:dcond2}
\end{equation}

\subsection{Second order terms}
At second order in the deformation the conditions (\ref{eq:cond1}) and (\ref{eq:cond2}) read
\begin{equation}
(P_+)_A{}^CE_C{}^N\Theta_N{}^M\partial_MK^{(-)}_B=0\,,
\qquad
E_A{}^N\Theta_N{}^M\partial_MK^{(-)A}+K^AK^{(-)}_A=0\,.
\end{equation}
We need to evaluate
\begin{align}
E_A{}^N\Theta_N{}^M\partial_MK_B
=&
E_A{}^NE_B{}^L\Theta_N{}^M\partial_MK_L
+E_A{}^N\Theta_N{}^M\partial_ME_B{}^LK_L
\nonumber\\
=&
R^{rs}E_A{}^NE_B{}^Mk_{rN}\left(k_s^L\partial_LK_M-K^L\partial_Lk_{sM}\right)\,,
\end{align}
where we used the fact that $\Theta^{mn}=k^m_rk^n_sR^{rs}$ and the isometry of the generalized vielbein, i.e. its generalized Lie derivative (\ref{eq:genLie}) along $k_r$ vanishes, in the second step. Now we use the form of $K^m$ in (\ref{eq:Km}) and the algebra of the Killing vectors to reduce this to
\begin{equation}
E_A{}^N\Theta_N{}^M\partial_MK_B
=
-\tfrac12k_{rA}k_{wB}R^{rs}f_{sv}{}^wR^{tu}f_{tu}{}^v\,.
\end{equation}
Using the Jacobi identity and the (modified) classical YB equation this expression can be seen to be symmetric in the indices $A$ and $B$. The second order conditions now become
\begin{equation}
(G-B)_{an}k^n_r(G+B)_{bm}k^m_wR^{rs}f_{sv}{}^wR^{tu}f_{tu}{}^v=0\,,\qquad K^2+\tilde K^2=0\,.
\label{eq:2nd-cond}
\end{equation}
The first condition can be expressed as
\begin{equation}
k^n_rk^m_wR^{rs}f_{sv}{}^wR^{tu}f_{tu}{}^v=v_+^mv_+^n+v_-^mv_-^n\,,
\end{equation}
where $v_\pm$ are zero-eigenvectors of $G\pm B$, i.e. $(G\pm B)v_\pm=0$. When $G\pm B$ is degenerate precisely one such vector $v_\pm$ exists (up to rescaling). When $G\pm B$ is non-degenerate, for example if $B$ vanishes, then the RHS is zero and we get the condition
\begin{equation}
k^n_rk^m_wR^{rs}f_{sv}{}^wR^{tu}f_{tu}{}^v=0\,.
\end{equation}
This condition is very strong and in fact it seems to imply the unimodularity condition, at least for deformations of symmetric spaces. In that case the condition becomes (see footnote \ref{foot:sym})
\begin{equation}
(\hat P\mathrm{Ad}_g)^a{}_r(\hat P\mathrm{Ad}_g)^b{}_wR^{rs}f_{sv}{}^wR^{tu}f_{tu}{}^v=0\,,
\end{equation}
and taking $g=e^{\epsilon^aP_a}$ and expanding in $\epsilon$ this leads to
\begin{equation}
R^{rs}f_{sv}{}^wR^{tu}f_{tu}{}^v=0\,.
\end{equation}
It is easy to see from the form of the Drinfeld-Jimbo $R$-matrix that this rules out the inhomogeneous deformations. For the homogeneous deformations $R$ is invertible on the subalgebra where it is defined and this condition is equivalent to the condition that the distinguished Lie algebra element $R^{rs}f_{rs}{}^tT_t$ must lie in the center of the algebra.\footnote{It also implies that the algebra can be constructed as a so-called symplectic double extension of a lower-dimensional symplectic, or quasi-Frobenius, Lie algebra \cite{Medina2007}. The question is then if the symplectic double extension of a unimodular Lie algebra is always unimodular, in which case this condition would imply unimodularity.} While we have not found a general proof that this implies unimodularity one can easily verify that this is true for $R$-matrices of rank$<8$. In the rank 2 case this is trivial to see. For rank 4  the relevant algebras are classified in \cite{ovando2006four} and it is easy to check that only unimodular examples satisfy the condition. For rank 6 the relevant algebras are classified in \cite{Campoamor:2005} (nilpotent algebras are automatically unimodular) and again only unimodular ones satisfy the condition. In addition we note that for $AdS_5$, corresponding to the isometry group $SO(2,4)$, the maximum rank of $R$ is 8 \cite{Borsato:2016ose}, however it is easy to see that the 8-dimensional algebras in question have a trivial center and can therefore not lead to any exception to the unimodularity condition. This rules out non-unimodular deformations of $AdS_n$ with $n\leq5$ if $G+B$ is invertible.

Therefore we conclude that for deformations of symmetric spaces non-unimodular $R$-matrices can lead to one-loop Weyl invariant $\sigma$-models only if $G\pm B$ of the undeformed model is degenerate (with the caveat that we checked this only up to rank 6). In that case they must satisfy (\ref{eq:2nd-cond}) as well as the conditions we found at first order, namely that $\tilde K=i_KB$ generates isometries of $G,H,\Phi$ and equation (\ref{eq:dK}).\footnote{For general inhomogeneous deformations with WZ-term a more careful analysis, where the condition of invariance of $B$ is dropped, is required.} Examples of such backgrounds were found in \cite{Sakamoto:2018krs,Borsato:2018spz}.

We will now turn to the question of what happens at two loops, i.e. including the first $\alpha'$-correction to the (super)gravity equations of motion. We will find that the conditions at two loops as actually weaker. We will only need to satisfy the conditions we found at first order in the deformation to solve also the two-loop equations.

\section{Two-loop Weyl invariance}\label{sec:2loop}
Here we will show that the $\alpha'$-correction to the equations of motion can be cast in a form that is manifestly invariant under non-unimodular YB deformations satisfying the one-loop Weyl invariance conditions of the previous section. In fact our calculation will be more general. We will assume only that the following remain invariant under the transformation in question
\begin{equation}
\mathcal F_{ABC}\,,\qquad\partial_{A_1}\cdots\partial_{A_n}(\mbox{anything invariant})\,,\qquad\mathcal F^A\mathcal F^{(\pm)}_{ABC}\,,\qquad\mathcal F^{(\pm)}_A\partial^A(\mbox{anything invariant})\,.
\label{eq:invariant}
\end{equation}
However, $\mathcal F_A$ and its derivatives need not be invariant. As we have seen this is true for any YB deformation that is one-loop Weyl invariant (\ref{eq:dcond1}), (\ref{eq:dcond2}) (it is trivially true for unimodular deformations since in that case also $\mathcal F_A$ is invariant under the deformation).

To get the equations of motion at order $\alpha'$ we must vary the corrected action (\ref{eq:S1}) using the expressions for the variations of the fluxes in (\ref{eq:deltaF}). The variation with respect to the generalized dilaton is easy and gives just the vanishing of the Lagrangian itself
\begin{equation}
\mathcal R+a\mathcal R^{(-)}+b\mathcal R^{(+)}=0\,.
\end{equation}
In the following we will set $b=0$ to simplify the calculations. In the end our results will apply also for $b\neq0$. Displaying only the order $\alpha'$-terms that are not trivially  invariant under the YB deformation we have from (\ref{eq:R-})
\begin{equation}
\mathcal R^{(-)}=
-2\partial^A\left[\mathcal F^B\tr\big(\mathcal F^{(-)}_A\mathcal F^{(-)}_B\big)\right]
+\mathcal F^A\mathcal F^B\tr\big(\mathcal F^{(-)}_A\mathcal F^{(-)}_B\big)
+\ldots
\end{equation}
where the ellipsis denotes terms involving only $\mathcal F_{ABC}$, which are trivially invariant. Using the invariance of the expressions in (\ref{eq:invariant}) we see that the RHS is invariant. Therefore the dilaton equation remains satisfied to order $\alpha'$ for such deformations.

Varying the action (\ref{eq:S1}) with respect to the generalized vielbein using (\ref{eq:deltaF}) the terms involving $\mathcal F_A$, i.e. the first two terms, in $\mathcal R^{(-)}$ (\ref{eq:R-}) give the following contributions to the equations of motion
\begin{align}
&2(\partial^C-\mathcal F^C)\left[(\partial^D\mathcal F^{(+)}_A+\partial^{(+)}_A\mathcal F^D)\mathcal F^{(-)}_{DCB}\right]
-(\partial_C-\mathcal F_C)\left[(\partial^D\mathcal F^{(+)C}+\partial^{(+)C}\mathcal F^D)\mathcal F^{(-)}_{DAB}\right]
\nonumber\\
&{}
+(\partial^{(-)}_A\mathcal F^C-\partial^C\mathcal F^{(+)}_A)\tr\big(\mathcal F^{(-)}_C\mathcal F^{(-)}_B\big)
-(\partial^C\mathcal F^{(+)}_A+\partial^{(+)}_A\mathcal F^C)\tr\big(\mathcal F^{(-)}_C\mathcal F^{(--)}_B\big)
\nonumber\\
&{}
+2(\partial^C\mathcal F^D+\partial^D\mathcal F^C)\mathcal F^{(+)E}{}_{CA}\mathcal F^{(-)}_{DEB}
-(A\leftrightarrow B)
+\ldots\,,
\label{eq:contr1}
\end{align}
where we suppress terms that are manifestly invariant, i.e. constructed form the invariant combinations in (\ref{eq:invariant}). The variation of the $\mathcal R_{AB}^2$-term in (\ref{eq:R-}) gives rise to the terms
\begin{align}
&4\partial^C\left[\partial^{(+)}_B\mathcal F^D\mathcal F^{(-)}_{DCA}\right]
+4\mathcal F^C(\partial^D-\mathcal F^D)\bar{\mathcal R}^{(-)}_{DBCA}
-4\partial^C\left[\mathcal F^D\mathcal F^{(++)}_{DBE}\mathcal F^{(-)E}{}_{CA}\right]
\nonumber\\
&{}
+2\partial^{(+)}_A\mathcal F^C\tr\big(\mathcal F^{(-)}_C\mathcal F^{(--)}_B\big)
-4\partial^{(+)C}\mathcal F^D\mathcal F^{(+)E}{}_{CA}\mathcal F^{(-)}_{DEB}
+2\mathcal F^C\mathcal F^{(++)}_{ACD}\tr\big(\mathcal F^{(-)D}\mathcal F^{(--)}_B\big)
\nonumber\\
&{}
+4\mathcal F_C\mathcal F^{(++)CEF}\mathcal F^{(+)D}{}_{EA}\mathcal F^{(-)}_{FDB}
-2\mathcal F^C\mathcal F^{(++)DE}{}_A\bar{\mathcal R}^{(-)}_{DECB}
-4\mathcal F^C\mathcal F^{(-)DE}{}_B\bar{\mathcal R}^{(-)}_{ADEC}
\nonumber\\
&{}
+4\mathcal F_C\mathcal F^{(-)DEC}\bar{\mathcal R}^{(-)}_{ADEB}
-(A\leftrightarrow B)
+\ldots\,,
\label{eq:contr2}
\end{align}
where we have noted that using the definition (\ref{eq:RAB}) we have
\begin{equation}
\mathcal F^A\bar{\mathcal R}^{(-)}_{ABCD}
=
\partial^{(+)}_B\mathcal F^A\mathcal F^{(-)}_{ACD}
-\mathcal F^A\mathcal F^{(++)}_{ABE}\mathcal F^{(-)E}{}_{CD}
+\ldots\,.
\end{equation}
Finally the variation of the $\mathcal C_{ABC}$-term in (\ref{eq:R-}) gives
\begin{align}
&\partial_A\mathcal F^C\tr\big(\mathcal F^{(-)}_B\mathcal F^{(-)}_C\big)
-\mathcal F^{(+)}_C\mathcal F^{CDE}\mathcal R^{(-)}_{DEAB}
-2\mathcal F^C\mathcal F^{(++)DE}{}_B\bar{\mathcal R}^{(-)}_{DECA}
-4\mathcal F^C\mathcal F^{(+)DE}{}_B\mathcal R^{(-)}_{DECA}
\nonumber\\
&{}
-\partial_C\left[\mathcal F_D\mathcal F^{(++)CDE}\mathcal F^{(-)}_{EAB}\right]
+2(\partial^C-\mathcal F^C)\left[\mathcal F_D\mathcal F^{(++)DE}{}_A\mathcal F^{(-)}_{ECB}\right]
\nonumber\\
&{}
+2\mathcal F_C\partial^D\mathcal F^{(+)}_{DBE}\mathcal F^{(-)EC}{}_A
+2\mathcal F_C\partial^D\mathcal F^{(++)}_{DBE}\mathcal F^{(-)EC}{}_A
-\mathcal F^{(+)}_C\partial_D\mathcal F^{CDE}\mathcal F^{(-)}_{EAB}
\nonumber\\
&{}
-\mathcal F_C\mathcal F_D\mathcal F^{(+)CDE}\mathcal F^{(-)}_{EAB}
-2\mathcal F^C\mathcal F^D\mathcal F^{(+)}_{CEA}\mathcal F^{(-)E}{}_{DB}
+2\mathcal F^C\mathcal F^{(+)}_{ACD}\tr\big(\mathcal F^{(-)D}\mathcal F^{(-)}_B\big)
\nonumber\\
&{}
+\mathcal F^C\mathcal F^{(++)}_{ACD}\tr\big(\mathcal F^{(-)D}\mathcal F^{(-)}_B\big)
-\mathcal F^C\mathcal F^{(++)}_{ACD}\tr\big(\mathcal F^{(-)D}\mathcal F^{(--)}_B\big)
+2\mathcal F_C\mathcal F^{(++)CDE}\mathcal F_{DA}{}^F\mathcal F^{(-)}_{EBF}
\nonumber\\
&{}
+\mathcal F_C\mathcal F^{(+)EFC}\mathcal F^{(+)}_{EFD}\mathcal F^{(-)D}{}_{AB}
+2\mathcal F^C\mathcal F^{(+)}_{EFB}\mathcal F^{(+)EFD}\mathcal F^{(-)}_{DCA}
-(A\leftrightarrow B)
+\ldots\,.
\label{eq:contr3}
\end{align}
Now we need to add together these three potentially non-invariant contributions to the equations of motion. 

Using the Bianchi identity for $\mathcal F_A$ (\ref{eq:Bianchi}) and noting also that the second term in (\ref{eq:contr1}) can be written
\begin{align}
2\partial^{(+)}_C\left(\partial^{(C}\mathcal F^{D)}\mathcal F^{(-)}_{DAB}\right)
=&
\partial^C\left(\mathcal F^D\mathcal F^{(++)}_{CDE}\mathcal F^{(-)E}{}_{AB}-2\partial^{(-)}_C\mathcal F^D\mathcal F^{(-)}_{DAB}\right)
+2\mathcal F^C\partial_C\mathcal F^D\mathcal F^{(-)}_{DAB}
+\ldots
\nonumber\\
=&
\partial^C\left(\mathcal F^D[\mathcal F^{(++)}_{CDE}+2\mathcal F^{(+)}_{CDE}]\mathcal F^{(-)E}{}_{AB}\right)
+2\mathcal F^C\partial_C\mathcal F^D\mathcal F^{(-)}_{DAB}
+\ldots
\end{align}
we find, after a bit of algebra, that all terms involving only $\mathcal F^{(+)}_A$ can be eliminated leaving the terms
\begin{align}
&{}8\mathcal F^C\partial^D\partial^{(+)}_{[A}\mathcal F^{(-)}_{D]CB}
-4\mathcal F^C\partial^D[\mathcal F^{(-)}_{AC}{}^E\mathcal F^{(-)}_{DEB}]
-4\mathcal F^C\partial^D[\mathcal F^{(-)}_{AB}{}^E\mathcal F^{(-)}_{DCE}]
-4\mathcal F^C\mathcal F^{(++)}_{AD}{}^E\partial^D\mathcal F^{(-)}_{ECB}
\nonumber\\
&{}
-8\mathcal F^C\mathcal F^{(-)DE}{}_B\partial^{(+)}_{[A}\mathcal F^{(-)}_{D]EC}
+8\mathcal F_C\mathcal F^{(-)DEC}\partial^{(+)}_{[A}\mathcal F^{(-)}_{D]EB}
+4\mathcal F^C\partial^D[\mathcal F^{(+)}_{DEA}\mathcal F^{(-)E}{}_{CB}]
\nonumber\\
&{}
-8\mathcal F^C\mathcal F^D\partial^{(+)}_{[A}\mathcal F^{(-)}_{D]CB}
-4\mathcal F^C\partial^{(+)}_A\mathcal F^D\mathcal F^{(-)}_{DCB}
-4\mathcal F^C\mathcal F^D\mathcal F^{(+)}_{CEA}\mathcal F^{(-)E}{}_{DB}
+4\mathcal F^C\mathcal F^D\mathcal F^{(-)}_{AC}{}^E\mathcal F^{(-)}_{DEB}
\nonumber\\
&{}
+4\mathcal F_C\mathcal F_D\mathcal F^{(-)}_{ABE}\mathcal F^{(-)CDE}
-8\mathcal F_C\mathcal F^{(-)DEC}\mathcal F^{(++)}_{AD}{}^F\mathcal F^{(-)}_{FEB}
-8\mathcal F_C\mathcal F^{(-)DEC}\mathcal F^{(-)}_{AE}{}^F\mathcal F^{(-)}_{DFB}
\nonumber\\
&{}
-4\mathcal F^C\mathcal F^{(-)}_{DEC}\mathcal F^{(-)DEF}\mathcal F^{(-)}_{ABF}
-4\mathcal F^C\mathcal F^{(-)}_{AFC}\mathcal F^{(-)}_{DEB}\mathcal F^{(-)DEF}
-4\mathcal F^C\mathcal F^{(-)}_{FCB}\mathcal F^{(+)}_{DEA}\mathcal F^{(+)DEF}
\nonumber\\
&{}
-2\mathcal F_C\left(\partial^{(-)C}\mathcal F^{(+)D}+(\partial_E-\mathcal F_E)\mathcal F^{(+)CDE}-\mathcal F^{(-)EFC}\mathcal F^{(+)}_{FE}{}^D\right)\mathcal F^{(-)}_{DAB}
\nonumber\\
&{}
+2\left(\partial^{(+)}_C\mathcal F^{(-)}_A-\mathcal F^E\mathcal F^{(-)}_{CAE}\right)\tr\big(\mathcal F^{(-)}_B\mathcal F^{(-)C}\big)
-(A\leftrightarrow B)
+\ldots
\end{align}
The last two terms drop out using the lowest order equations of motion (\ref{eq:eom}). We now rewrite the first term as
\begin{align}
\lefteqn{
8\mathcal F^D\partial^C\partial^{(-)}_{[B}\mathcal F^{(+)}_{D]AC}
-8\mathcal F^D\partial^C\left(\partial^{(+)}_{[A}\mathcal F^{(-)}_{C]BD}+\partial^{(-)}_{[B}\mathcal F^{(+)}_{D]AC}\right)
}
\nonumber\\
=&
-8\mathcal F^D\partial^{(-)}_{[D}\left(\partial^{(-)}_{B]}\mathcal F^{(+)}_A+(\partial^C-\mathcal F^C)\mathcal F^{(+)}_{B]AC}-\mathcal F^{(-)}_{EFB}\mathcal F^{(+)FE}{}_A\right)
-4\mathcal F^D\mathcal F^{(--)}_{BDE}\partial^E\mathcal F^{(+)}_A
\nonumber\\
&{}
-4\mathcal F^D\mathcal F^{(-)}_{EBD}\partial^E\mathcal F^{(+)}_A
+8\mathcal F^D\mathcal F^{(+)}_{[D|AC|}\partial^{(-)}_{B]}\mathcal F^{(+)C}
-8\mathcal F^C\mathcal F^D\partial^{(+)}_{[A}\mathcal F^{(-)}_{C]BD}
+4\mathcal F_D\mathcal F^{(+)}_{FEA}\partial^{(-)}_B\mathcal F^{(-)EFD}
\nonumber\\
&{}
+8\mathcal F_D\mathcal F^{(-)EFD}\partial^{(+)}_{[A}\mathcal F^{(-)}_{E]BF}
+4\mathcal F^D\mathcal F^{(+)}_{BC}{}^E\partial^C\mathcal F^{(+)}_{DAE}
-4\mathcal F^D\mathcal F^{(-)E}{}_{CB}\partial^C\mathcal F^{(+)}_{DAE}
\nonumber\\
&{}
-8\mathcal F^D(\partial^C-\mathcal F^C)\left(\partial^{(+)}_{[A}\mathcal F^{(-)}_{C]BD}+\partial^{(-)}_{[B}\mathcal F^{(+)}_{D]AC}\right)
-8\mathcal F_E\mathcal F^{(-)CDE}\left(\partial^{(+)}_{[A}\mathcal F^{(-)}_{C]BD}+\partial^{(-)}_{[B}\mathcal F^{(+)}_{D]AC}\right)
+\ldots
\end{align}
The first term vanishes by the lowest order equations of motion (\ref{eq:eom}). In the last two terms we can use the Bianchi identity for $\mathcal F_{ABC}$ (\ref{eq:Bianchi}), which implies in particular that
\begin{align}
2\partial^{(+)}_{[A}\mathcal F^{(-)}_{C]BD}+2\partial^{(-)}_{[B}\mathcal F^{(+)}_{D]AC}
=&
\mathcal F^{(-)}_{AB}{}^E\mathcal F^{(-)}_{CED}
-\mathcal F^{(+)}_{BA}{}^E\mathcal F^{(+)}_{DCE}
+\mathcal F^{(-)}_{AD}{}^E\mathcal F^{(-)}_{CBE}
\nonumber\\
&{}
-\mathcal F^{(+)}_{DA}{}^E\mathcal F^{(+)}_{BEC}
+\mathcal F^{(++)}_{AC}{}^E\mathcal F^{(-)}_{EBD}
+\mathcal F^{(+)E}{}_{AC}\mathcal F^{(--)}_{BDE}\,.
\end{align}
After a bit of algebra we are left with
\begin{align}
&{}8\mathcal F^D\mathcal F^{(+)}_{[D|A|}{}^C\left(\partial^{(-)}_{B]}\mathcal F^{(+)}_C+(\partial^E-\mathcal F^E)\mathcal F^{(+)}_{B]CE}-\mathcal F^{(-)EF}{}_{B]}\mathcal F^{(+)}_{FEC}\right)
\nonumber\\
&{}
-4\mathcal F^D\mathcal F^{(--)}_{BD}{}^E\left(\partial^{(-)}_E\mathcal F^{(+)}_A+(\partial^C-\mathcal F^C)\mathcal F^{(+)}_{EAC}-\mathcal F^{(-)}_{CFE}\mathcal F^{(+)FC}{}_A\right)
\nonumber\\
&{}
-4\mathcal F_E\mathcal F^{(-)CDE}\left(
\mathcal F^{(-)}_{AD}{}^F\mathcal F^{(-)}_{CFB}
-\mathcal F^{(+)}_{DA}{}^F\mathcal F^{(+)}_{BFC}
+\mathcal F^{(++)}_{AC}{}^F\mathcal F^{(-)}_{FDB}
+\mathcal F^{(+)F}{}_{AC}\mathcal F^{(--)}_{BDF}
\right)
\nonumber\\
&{}
-8\mathcal F^D\mathcal F^{(-)CE}{}_B\left(\partial^{(+)}_{[A}\mathcal F^{(-)}_{C]ED}+\partial^{(-)}_{[E}\mathcal F^{(+)}_{D]AC}\right)
+4\mathcal F^D\mathcal F^{(--)}_{BDE}\mathcal F^{(-)CEF}\mathcal F^{(+)}_{FCA}
\nonumber\\
&{}
-4\mathcal F^C\mathcal F^{(-)}_{AFC}\mathcal F^{(-)}_{DEB}\mathcal F^{(-)DEF}
+4\mathcal F^C\mathcal F^{(+)DE}{}_A\left(\partial_E\mathcal F^{(--)}_{BCD}-\partial^{(-)}_B\mathcal F^{(-)}_{ECD}+\partial_D\mathcal F^{(-)}_{ECB}\right)
\nonumber\\
&{}
-4\mathcal F^C\mathcal F^{(-)}_{FCB}\mathcal F^{(+)}_{DEA}\mathcal F^{(+)DEF}
-(A\leftrightarrow B)
+\ldots\,.
\end{align}
The first two terms vanish by the lowest order equations of motion and the remaining terms cancel using the Bianchi identity for $\mathcal F_{ABC}$. This completes the proof that the $\alpha'$-correction to the equations of motion can be cast in a manifestly invariant form provided that the expressions in (\ref{eq:invariant}) are invariant. In particular this implies that if a YB deformation preserves Weyl invariance at one-loop it also preserves it at two loops.

\section{Conclusions}
We have analyzed the conditions for a YB deformation of the bosonic/heterotic string sigma-model to be Weyl-invariant at one loop, i.e. for the corresponding background to be a (super)gravity solution. When $(G+B)_{mn}$ of the undeformed background is invertible one finds no solution in the inhomogeneous case (although our analysis for the YB model with WZ-term is not quite complete). For a homogeneous deformation of a symmetric space one finds that the distinguished Lie algebra element $R^{rs}f_{rs}{}^tT_t$ must belong to the center of the algebra. We showed that, at least for rank $R<8$, this in fact implies the usual unimodularity condition $R^{rs}f_{rs}{}^t=0$ of \cite{Borsato:2016ose}. When $(G+B)_{mn}$ of the undeformed background is non-invertible instead the unimodularity condition is replaced by the weaker conditions (\ref{eq:dK}), (\ref{eq:2nd-cond}) together with the condition that $\tilde K=i_KB$ generate isometries of the undeformed background $G,H,\Phi$. This is consistent with what has been seen in specific examples \cite{Sakamoto:2018krs,Borsato:2018spz} and the conditions we find agree with those coming from an analysis of generalized supergravity, see appendix E of \cite{Borsato:2018spz}, when specifying to YB deformations. We have also seen that when these conditions are satisfied the deformation in fact preserves Weyl-invariance at least to two loops, i.e. the background solves the low-energy effective string equations including the first $\alpha'$-correction.

Interestingly, while in the case of unimodular deformations the fact that the two-loop equations are satisfied is trivial in the doubled formulation we are using, this is not the case for non-unimodular ones due to the shift of $\mathcal F_A$ by the generalized Killing vector $K_A$. In fact it took quite a bit of work to show that the equations of motion can be cast in a form where it is easy to see that they are invariant under the deformation. It would be interesting to understand if one can improve the formulation so that the invariance is manifest also in the non-unimodular case and, if so, what this implies for the structure of higher-derivative corrections. Perhaps the natural starting point to analyzing this question is the gauged version of DFT \cite{Grana:2012rr}.

For unimodular YB deformations the first $\alpha'$-correction to the deformed background was derived in \cite{Borsato:2020bqo}, also by using the doubled formulation. The same correction is valid also for the non-unimodular examples discussed here.

It would be interesting to extend our analysis to the general case of inhomogeneous YB deformation with WZ-term by relaxing the requirement that $B$ is invariant under the isometries. The conditions must become essentially the same in that case since they are mostly fixed by the generalized supergravity analysis. It would be interesting to understand if there exist any non-unimodular Weyl-invariant examples in that case. It seems unlikely to be the case since $R$ is much more constrained than in the homogeneous case.

Finally, it would be interesting to extend the present analysis to the case of Poisson-Lie T-duality, for which the first $\alpha'$-correction was recently found \cite{Borsato:2020wwk,Hassler:2020tvz,Codina:2020yma} using essentially the same approach as for YB.

\section*{Acknowledgements}
We thank R. Borsato, D. Marqu\'es and S. van Tongeren for interesting discussions and R. Borsato for comments on the manuscript. The work of LW is supported by the grant ``Integrable Deformations'' (GA20-04800S) from the Czech Science Foundation (GA\v CR).

\bibliographystyle{nb}
\bibliography{biblio}{}

\begin{thebibliography}{10}
\ifx\href\asklfhas\newcommand{\href}[2]{#2}\fi
\ifx\arxivref\asklfhas\newcommand{\arxivref}[2]{\href{http://arxiv.org/abs/#1}{#2}}\fi
\ifx\doiref\asklfhas\newcommand{\doiref}[2]{\href{http://dx.doi.org/#1}{#2}}\fi
\raggedright
\small
\parskip 0pt

\bibitem{Klimcik:2002zj}
C.~Klimcik,
\textit{``{Yang-Baxter sigma models and dS/AdS T duality}''},
\textsf{\doiref{10.1088/1126-6708/2002/12/051}{JHEP~0212,~051~(2002)}},
\texttt{\arxivref{hep-th/0210095}{hep-th/0210095}}.

\bibitem{Klimcik:2008eq}
C.~Klimcik,
\textit{``{On integrability of the Yang-Baxter sigma-model}''},
\textsf{\doiref{10.1063/1.3116242}{J.Math.Phys.~50,~043508~(2009)}},
\texttt{\arxivref{0802.3518}{arxiv:0802.3518}}.

\bibitem{Delduc:2013fga}
F.~Delduc, M.~Magro and B.~Vicedo,
\textit{``{On classical $q$-deformations of integrable sigma-models}''},
\textsf{\doiref{10.1007/JHEP11(2013)192}{JHEP~1311,~192~(2013)}},
\texttt{\arxivref{1308.3581}{arxiv:1308.3581}}.

\bibitem{Delduc:2013qra}
F.~Delduc, M.~Magro and B.~Vicedo,
\textit{``{An integrable deformation of the AdS$_5 \times$S$^5$ superstring
  action}''},
\textsf{\doiref{10.1103/PhysRevLett.112.051601}{Phys.Rev.Lett.~112,~051601~(2014)}},
\texttt{\arxivref{1309.5850}{arxiv:1309.5850}}.

\bibitem{Kawaguchi:2014qwa}
I.~Kawaguchi, T.~Matsumoto and K.~Yoshida,
\textit{``{Jordanian deformations of the $AdS_5 x S^5$ superstring}''},
\textsf{\doiref{10.1007/JHEP04(2014)153}{JHEP~1404,~153~(2014)}},
\texttt{\arxivref{1401.4855}{arxiv:1401.4855}}.

\bibitem{Hoare:2016wsk}
B.~Hoare and A.~A.~Tseytlin,
\textit{``{Homogeneous Yang-Baxter deformations as non-abelian duals of the
  AdS$_5$ sigma-model}''},
\textsf{\doiref{10.1088/1751-8113/49/49/494001}{J.~Phys.~A49,~494001~(2016)}},
\texttt{\arxivref{1609.02550}{arxiv:1609.02550}}.

\bibitem{Borsato:2016pas}
R.~Borsato and L.~Wulff,
\textit{``{Integrable Deformations of $T$-Dual $\sigma$ Models}''},
\textsf{\doiref{10.1103/PhysRevLett.117.251602}{Phys.~Rev.~Lett.~117,~251602~(2016)}},
\texttt{\arxivref{1609.09834}{arxiv:1609.09834}}.

\bibitem{Borsato:2017qsx}
R.~Borsato and L.~Wulff,
\textit{``{On non-abelian T-duality and deformations of supercoset string
  sigma-models}''},
\textsf{\doiref{10.1007/JHEP10(2017)024}{JHEP~1710,~024~(2017)}},
\texttt{\arxivref{1706.10169}{arxiv:1706.10169}}.

\bibitem{Arutyunov:2015qva}
G.~Arutyunov, R.~Borsato and S.~Frolov,
\textit{``{Puzzles of $\eta$-deformed AdS$_5 \times$ S$^5$}''},
\textsf{\doiref{10.1007/JHEP12(2015)049}{JHEP~1512,~049~(2015)}},
\texttt{\arxivref{1507.04239}{arxiv:1507.04239}}.

\bibitem{Arutyunov:2015mqj}
G.~Arutyunov, S.~Frolov, B.~Hoare, R.~Roiban and A.~A.~Tseytlin,
\textit{``{Scale invariance of the $\eta$-deformed $AdS_5\times S^5$
  superstring, T-duality and modified type II equations}''},
\textsf{\doiref{10.1016/j.nuclphysb.2015.12.012}{Nucl.~Phys.~B903,~262~(2016)}},
\texttt{\arxivref{1511.05795}{arxiv:1511.05795}}.

\bibitem{Wulff:2016tju}
L.~Wulff and A.~A.~Tseytlin,
\textit{``{Kappa-symmetry of superstring sigma model and generalized 10d
  supergravity equations}''},
\textsf{\doiref{10.1007/JHEP06(2016)174}{JHEP~1606,~174~(2016)}},
\texttt{\arxivref{1605.04884}{arxiv:1605.04884}}.

\bibitem{Borsato:2016ose}
R.~Borsato and L.~Wulff,
\textit{``{Target space supergeometry of $\eta$ and $\lambda$-deformed
  strings}''},
\textsf{\doiref{10.1007/JHEP10(2016)045}{JHEP~1610,~045~(2016)}},
\texttt{\arxivref{1608.03570}{arxiv:1608.03570}}.

\bibitem{Borsato:2018idb}
R.~Borsato and L.~Wulff,
\textit{``{Non-abelian T-duality and Yang-Baxter deformations of Green-Schwarz
  strings}''},
\textsf{\doiref{10.1007/JHEP08(2018)027}{JHEP~1808,~027~(2018)}},
\texttt{\arxivref{1806.04083}{arxiv:1806.04083}}.

\bibitem{Osten:2016dvf}
D.~Osten and S.~J.~van~Tongeren,
\textit{``{Abelian Yang–Baxter deformations and TsT transformations}''},
\textsf{\doiref{10.1016/j.nuclphysb.2016.12.007}{Nucl.~Phys.~B915,~184~(2017)}},
\texttt{\arxivref{1608.08504}{arxiv:1608.08504}}.

\bibitem{Sakamoto:2018krs}
J.-I.~Sakamoto and Y.~Sakatani,
\textit{``{Local $\beta$-deformations and Yang-Baxter sigma model}''},
\textsf{\doiref{10.1007/JHEP06(2018)147}{JHEP~1806,~147~(2018)}},
\texttt{\arxivref{1803.05903}{arxiv:1803.05903}}.

\bibitem{Borsato:2018spz}
R.~Borsato and L.~Wulff,
\textit{``{Marginal deformations of WZW models and the classical Yang-Baxter
  equation}''},
\textsf{\doiref{10.1088/1751-8121/ab1b9c}{J.~Phys.~A52,~225401~(2019)}},
\texttt{\arxivref{1812.07287}{arxiv:1812.07287}}.

\bibitem{Delduc:2014uaa}
F.~Delduc, M.~Magro and B.~Vicedo,
\textit{``{Integrable double deformation of the principal chiral model}''},
\textsf{\doiref{10.1016/j.nuclphysb.2014.12.018}{Nucl.~Phys.~B~891,~312~(2015)}},
\texttt{\arxivref{1410.8066}{arxiv:1410.8066}}.

\bibitem{Geissbuhler:2013uka}
D.~Geissbuhler, D.~Marques, C.~Nunez and V.~Penas,
\textit{``{Exploring Double Field Theory}''},
\textsf{\doiref{10.1007/JHEP06(2013)101}{JHEP~1306,~101~(2013)}},
\texttt{\arxivref{1304.1472}{arxiv:1304.1472}}.

\bibitem{Borsato:2020bqo}
R.~Borsato, A.~Vilar~López and L.~Wulff,
\textit{``{The first $\alpha'$-correction to homogeneous Yang-Baxter
  deformations using $O(d,d)$}''},
\textsf{\doiref{10.1007/JHEP07(2020)103}{JHEP~2007,~103~(2020)}},
\texttt{\arxivref{2003.05867}{arxiv:2003.05867}}.

\bibitem{Grana:2012rr}
M.~Grana and D.~Marques,
\textit{``{Gauged Double Field Theory}''},
\textsf{\doiref{10.1007/JHEP04(2012)020}{JHEP~1204,~020~(2012)}},
\texttt{\arxivref{1201.2924}{arxiv:1201.2924}}.

\bibitem{Borsato:2019oip}
R.~Borsato and L.~Wulff,
\textit{``{Two-loop conformal invariance for Yang-Baxter deformed strings}''},
\textsf{\doiref{10.1007/JHEP03(2020)126}{JHEP~2003,~126~(2020)}},
\texttt{\arxivref{1910.02011}{arxiv:1910.02011}}.

\bibitem{Aldazabal:2013sca}
G.~Aldazabal, D.~Marques and C.~Nunez,
\textit{``{Double Field Theory: A Pedagogical Review}''},
\textsf{\doiref{10.1088/0264-9381/30/16/163001}{Class.~Quant.~Grav.~30,~163001~(2013)}},
\texttt{\arxivref{1305.1907}{arxiv:1305.1907}}.

\bibitem{Hohm:2013bwa}
O.~Hohm, D.~Lüst and B.~Zwiebach,
\textit{``{The Spacetime of Double Field Theory: Review, Remarks, and
  Outlook}''},
\textsf{\doiref{10.1002/prop.201300024}{Fortsch.~Phys.~61,~926~(2013)}},
\texttt{\arxivref{1309.2977}{arxiv:1309.2977}}.

\bibitem{Berman:2013eva}
D.~S.~Berman and D.~C.~Thompson,
\textit{``{Duality Symmetric String and M-Theory}''},
\textsf{\doiref{10.1016/j.physrep.2014.11.007}{Phys.~Rept.~566,~1~(2014)}},
\texttt{\arxivref{1306.2643}{arxiv:1306.2643}}.

\bibitem{Siegel:1993th}
W.~Siegel,
\textit{``{Superspace duality in low-energy superstrings}''},
\textsf{\doiref{10.1103/PhysRevD.48.2826}{Phys.~Rev.~D~48,~2826~(1993)}},
\texttt{\arxivref{hep-th/9305073}{hep-th/9305073}}.

\bibitem{Siegel:1993xq}
W.~Siegel,
\textit{``{Two vierbein formalism for string inspired axionic gravity}''},
\textsf{\doiref{10.1103/PhysRevD.47.5453}{Phys.~Rev.~D~47,~5453~(1993)}},
\texttt{\arxivref{hep-th/9302036}{hep-th/9302036}}.

\bibitem{Hohm:2010xe}
O.~Hohm and S.~K.~Kwak,
\textit{``{Frame-like Geometry of Double Field Theory}''},
\textsf{\doiref{10.1088/1751-8113/44/8/085404}{J.~Phys.~A~44,~085404~(2011)}},
\texttt{\arxivref{1011.4101}{arxiv:1011.4101}}.

\bibitem{Marques:2015vua}
D.~Marqu\'es and C.~A.~Nu\~nez,
\textit{``{T-duality and $\alpha'$-corrections}''},
\textsf{\doiref{10.1007/JHEP10(2015)084}{JHEP~1510,~084~(2015)}},
\texttt{\arxivref{1507.00652}{arxiv:1507.00652}}.

\bibitem{Baron:2017dvb}
W.~H.~Baron, J.~J.~Fernandez-Melgarejo, D.~Marques and C.~Nunez,
\textit{``{The Odd story of $\alpha'$-corrections}''},
\textsf{\doiref{10.1007/JHEP04(2017)078}{JHEP~1704,~078~(2017)}},
\texttt{\arxivref{1702.05489}{arxiv:1702.05489}}.

\bibitem{Araujo:2017jkb}
T.~Araujo, I.~Bakhmatov, E.~O.~Colg\'{a}in, J.~Sakamoto, M.~M.~Sheikh-Jabbari
  and K.~Yoshida,
\textit{``{Yang-Baxter $\sigma$-models, conformal twists, and noncommutative
  Yang-Mills theory}''},
\textsf{\doiref{10.1103/PhysRevD.95.105006}{Phys.~Rev.~D95,~105006~(2017)}},
\texttt{\arxivref{1702.02861}{arxiv:1702.02861}}.

\bibitem{Sakamoto:2017cpu}
J.-i.~Sakamoto, Y.~Sakatani and K.~Yoshida,
\textit{``{Homogeneous Yang-Baxter deformations as generalized
  diffeomorphisms}''},
\textsf{\doiref{10.1088/1751-8121/aa8896}{J.~Phys.~A50,~415401~(2017)}},
\texttt{\arxivref{1705.07116}{arxiv:1705.07116}}.

\bibitem{Bakhmatov:2018apn}
I.~Bakhmatov, E.~Ó~Colgáin, M.~Sheikh-Jabbari and H.~Yavartanoo,
\textit{``{Yang-Baxter Deformations Beyond Coset Spaces (a slick way to do
  TsT)}''},
\textsf{\doiref{10.1007/JHEP06(2018)161}{JHEP~1806,~161~(2018)}},
\texttt{\arxivref{1803.07498}{arxiv:1803.07498}}.

\bibitem{Araujo:2017enj}
T.~Araujo, E.~Ó~Colgáin, J.~Sakamoto, M.~Sheikh-Jabbari and K.~Yoshida,
\textit{``{$I$ in generalized supergravity}''},
\textsf{\doiref{10.1140/epjc/s10052-017-5316-5}{Eur.~Phys.~J.~C~77,~739~(2017)}},
\texttt{\arxivref{1708.03163}{arxiv:1708.03163}}.

\bibitem{Wulff:2018aku}
L.~Wulff,
\textit{``{Trivial solutions of generalized supergravity vs non-abelian
  T-duality anomaly}''},
\textsf{\doiref{10.1016/j.physletb.2018.04.025}{Phys.~Lett.~B781,~417~(2018)}},
\texttt{\arxivref{1803.07391}{arxiv:1803.07391}}.

\bibitem{Medina2007}
A.~Medina and P.~Revoy,
\textit{``Lattices in symplectic Lie groups''},
\textsf{Journal~of~Lie~Theory~17,~27–39~(2007)}.

\bibitem{ovando2006four}
G.~P.~Ovando,
\textit{``{Four dimensional symplectic Lie algebras}''},
\textsf{Beitr{\"a}ge~zur~Algebra~und~Geome~47,~419~(2006)},
\texttt{\arxivref{math/0407501}{math/0407501}}.

\bibitem{Campoamor:2005}
R.~Campoamor-Stursberg,
\textit{``Symplectic Forms on Six-dimensional Real Solvable Lie Algebras I''},
\textsf{\doiref{10.1142/S100538670900025X}{Algebra~Colloquium~16,~253~(2009)}},
\texttt{\arxivref{math/0507499}{math/0507499}}.

\bibitem{Borsato:2020wwk}
R.~Borsato and L.~Wulff,
\textit{``{Quantum correction to Poisson-Lie and non-abelian T-duality}''},
\texttt{\arxivref{2007.07902}{arxiv:2007.07902}}.

\bibitem{Hassler:2020tvz}
F.~Hassler and T.~Rochais,
\textit{``{$\alpha'$-corrected Poisson-Lie T-duality}''},
\texttt{\arxivref{2007.07897}{arxiv:2007.07897}}.

\bibitem{Codina:2020yma}
T.~Codina and D.~Marques,
\textit{``{Generalized Dualities and Higher Derivatives}''},
\texttt{\arxivref{2007.09494}{arxiv:2007.09494}}.

\end{thebibliography}

\end{document}